\documentclass[twocolumn]{aastex631}

\usepackage{amsmath}
\usepackage{hyperref}

\begin{document}

\title{Detection of a Transient Quasi-periodic Oscillation in $\gamma$-Rays from Blazar PKS 2255-282}

\author[0000-0002-5221-0822]{Ajay Sharma}
\altaffiliation{ajjjkhoj@gmail.com}
\affiliation{S. N. Bose National Centre for Basic Sciences, Block JD, Salt Lake, Kolkata 700106, India}

\author[0000-0001-7796-8907]{Anuvab Banerjee}
\altaffiliation{anuvab.ban@gmail.com}
\affiliation{Department of Physics and Astronomy, Clemson University, Kinard Lab of Physics, Clemson, SC 29634-0978, USA}

\author[0000-0002-9526-0870]{Avik Kumar das}
\altaffiliation{avikdas@iisermohali.ac.in}
\affiliation{Department of Physical Sciences, Indian Institute of Science Education and Research Mohali,\\
Knowledge City, Sector 81, SAS Nagar, Punjab 140306, India}

\author[0009-0004-0275-5201]{Avijit Mandal}
\altaffiliation{jabhi3795@gmail.com}
\affiliation{S. N. Bose National Centre for Basic Sciences, Block JD, Salt Lake, Kolkata 700106, India}

\author[0000-0003-1071-5854]{Debanjan Bose}
\altaffiliation{debanjan.tifr@gmail.com}
\affiliation{Department of Physics, Central University of Kashmir, Ganderbal, 191131, India}







\begin{abstract}
We conducted a comprehensive variability analysis of the blazar PKS 2255-282 using Fermi-LAT observations spanning over four years, from MJD 57783.5 to 59358.5. Our analysis revealed a transient quasi-periodic oscillation (QPO) with a period of 93$\pm$2.6 days. We employed a variety of Fourier-based methods, including the Lomb-Scargle Periodogram (LSP) and Weighted Wavelet Z-Transform (WWZ), as well as time domain analysis techniques such as Seasonal and Non-Seasonal Autoregressive Integrated Moving Average (ARIMA) models and the Stochastic modeling with Stochastically Driven Damped Harmonic Oscillator (SHO) models. Consistently, the QPO with a period of 93 days was detected across all methods used. The observed peak in LSP and time-averaged WWZ plots has a significance level of 4.06$\sigma$ and 3.96$\sigma$, respectively. To understand the source of flux modulations in the light curve, we explored various physical models. A plausible scenario involves the precession of the jet with a high Lorentz factor or the movement of a plasma blob along a helical trajectory within the relativistic jet.

\end{abstract}

\keywords{Active galactic nuclei (16) --- Jets (870) --- Gamma-rays (637)}

\section{Introduction} \label{sec:intro}
Blazars are the most powerful subclass of  active galactic nuclei (AGNs), emitting copiously in the entire electromagnetic waveband ranging from radio to high energy/very high energy gamma-rays. These sources are believed to harbor highly collimated relativistic jets pointing towards our line of sight (\citealt{Urry1995Sep}). Blazars are divided into two sub-classes based on the strength of the optical emission lines: Flat spectrum radio quasar (FSRQ) with broad and strong emission lines and BL Lacertae (BL Lac) object with weak or no emission lines. Broadband spectral energy distribution (SEDs) of blazars are characterized by a double-hump profile, the lower hump peaking at near-infrared (NIR) to X-ray waveband and the high energy hump peaking at MeV to GeV gamma-ray waveband. The low energy hump of blazar SED is believed to be produced by the synchrotron emission of the population of relativistic electrons/positrons (i.e. leptons) in the jet. However, the emission origin for the high energy hump remains a subject of debate within the blazar community. This hump can be explained by the leptonic and/or hadronic scenario. 
\\
Blazars exhibit stochastic flux variability across multiple wavelengths, with time scales ranging from minutes to years (\citealt{2014ApJ...786..143S}). However, several blazars' light curves show persistent (relative) and/or transient quasi-periodic variability (\citealt{Gierlinski2008, abdo2010gamma, King2013, 2014JApA...35..307G, 2014MNRAS.445L..16A, Ackermann2015} and reference therein). The gamma-ray emission of blazars originates in the relativistic jet. Therefore, QPOs reported in this band are crucial for understanding not only jet physics but also, indirectly, the particle acceleration mechanism. The continuous monitoring capability of Fermi-LAT (Large Area Telescope) has provided long-term temporal data for over 5000 gamma-ray sources (according to the 4FGL DR3 catalog), and utilizing this data, many strong QPOs in the gamma-ray band have been reported in the literature. The first gamma-ray QPO was reported by \cite{Ackermann2015} in blazar PG 1553+113. They observed a QPO period of 2.18$\pm$0.08 year with three complete cycles. It is believed to be a binary SMBH (Supermassive black-hole) system and thus a persistent QPO source \citep{Tavani2018Feb}. Since then, many more QPOs (both transient and persistent) have been reported in gamma-ray bright blazars (e.g., \citealt{Zhou2018Nov, Benkhali2020Feb, Penil2020Jun, Ren2023Apr, Das2023Jun, Prince2023Oct} and reference therein). A systematic periodicity search was performed \citep{Penil2020Jun} on a sample of 2274 AGNs detected by the Fermi-LAT instruments during its first 9 years of operation. This search led to the discovery of 11 persistent periodic candidates with a significant detection threshold ($\gtrsim$ 4$\sigma$). Later, \cite{penil2022evidence} reconducted the study on a sample of 24 most probable periodic blazars (sources were selected based on previous studies) using $\sim$12 years of Fermi-LAT data and reported that 5 of them exhibited persistent periodic behavior with high detection significance. \cite{Zhou2018Nov} first reported a significant transient QPO with a period of $\sim$ 34.5 days (significance level of $\sim$ 4.6$\sigma$ and persisted for 6 complete cycles) in the Fermi-LAT light curve of the blazar PKS 2247-131. They explained this QPO using a helical structure jet model, where the viewing angle of the emission zone changes periodically. Recently, \cite{Ren2023Apr} conducted a QPO study on 35 gamma-ray bright AGNs, revealing that 24 of them exhibited transient QPO-like behavior, with most persisting for 3 to 10 complete cycles. Only in one source, namely B2 1520+31, the transient QPO persisted for 17 complete cycles with a period of $\sim$ 39 days in the 7-day binned light curve. In few cases, they found multiple transient QPO-like features in the gamma-ray light curve (e.g., 3C 279). Based on previous studies, it appears that the occurrence of transient QPOs, lasting for nearly or less than 10 cycles and occurring on timescales ranging from months to years, are relatively common phenomena in $\gamma$-ray bright blazars. However, the exact physical origin of these QPOs remains ambiguous to the scientific community. \\
PKS 2255-282 is a FSRQ-type blazar with R.A. = 344.524875$^{\circ}$, Dec = -27.972556$^{\circ}$ \citep{Lanyi2010Mar} and located at a redshift z = 0.92584 \citep{Jones2009Oct}. This source was first identified as a gamma-ray emitter in 1997 by EGRET \citep{Macomb1999Mar} and has been under continuous monitoring by Fermi-LAT since 2008 (4FGL DR3 catalog name: 4FGL J2258.1-2759). \cite{Dutka2012Mar} reported the first high-activity phase of this source using Fermi-LAT in February 2012 with a daily averaged flux of (1.0$\pm$0.3)$\times$10$^{-6}$ ph cm$^{-2}$ s$^{-1}$. Since April 2017, this source has been exhibiting variability in gamma-ray band (see the light curve on \href{https://fermi.gsfc.nasa.gov/ssc/data/access/lat/LightCurveRepository/source.html?source_name=4FGL_J2258.1-2759}{Fermi LAT light curve repository}) and recently, on October 10 and 11, 2023, \cite{Zyl2023Oct} reported very high gamma-ray activity with a daily averaged flux of (2.1$\pm$0.2)$\times$10$^{-6}$ ph cm$^{-2}$ s$^{-1}$. This marks the highest reported daily flux ever observed by Fermi-LAT and thus far, no variability study has been conducted during the time duration of this phase. We are presenting a QPO study on this source in gamma-ray band for the first time. \\
The paper is organized as follows: In section \S\ref{sec:data}, we discuss the procedure of gamma-ray light-curve analysis. In section \S \ref{method}, we discuss different Fourier and non-Fourier-based methods used for transient QPO detection in the light curve. In section \S \ref{results}, we present the results of our QPO study, followed by a discussion on plausible physical scenarios for the observed transients QPO and conclusion in section \S \ref{discussion}.


\section{\textbf{Fermi-LAT DATA REDUCTION AND ANALYSIS}} \label{sec:data}
The study utilized data from Fermi-LAT. The LAT instrument is a pair conversion $\gamma$-ray detector with wide field of view which is about $\sim$2.4 sr, large effective area ($>$8000 $\rm{cm^2}$ at $\sim$1GeV), covering the energy range from $\sim$20MeV to $\sim$1 TeV and provide near-constant monitoring of the $\gamma$-ray sky in every 3 hours. A detailed description of Fermi-LAT can be found in \citealt{atwood2009large}.\par
In this study, the Fermi-LAT data was collected during the time span from February 2017 to June 2023 (MJD 57783.5 - 60114.5). For the purpose of analysis, a region of interest (ROI) was carefully chosen with 10$^\circ$ circular area centered at the source of interest PKS 2255-282 (RA: 344.52 and Dec: -27.97).\par
The \textit{Fermi} Science Tools, \texttt{FERMITOOLS} package\footnote{\url{https://fermi.gsfc.nasa.gov/ssc/data/analysis/documentation/}}, was used to analyze the $\gamma$-ray observations. The data analysis follows the standard criteria for point-source analysis. We employed the Pass 8 LAT database the Fermi Science Support Center provided. First, we have chosen the events belonging to the SOURCE class (evclass=128, evtype=3) using \texttt{GTSELECT} tool. To improve data quality, events with zenith angles over 90$^\circ$ excluded to avoid any contamination originating from the Earth's Limb. We applied the standard criteria with recommended filter expression \texttt{$\text{(DATA\_QUAL > 0) \&\& (LAT\_CONFIG == 1)}$}. The high-quality data with good time intervals (GTIs) were obtained using \texttt{GTMKTIME} tool. Other Fermi-LAT analysis tools like \texttt{GTLTCUBE} and \texttt{GTEXPSURE} were used to calculate the integrated livetime as a function of sky position, off-axis angle, and exposure, respectively. The galactic and extra-galactic diffuse background emissions were modeled using files \texttt{gll\_iem\_v07.fits}\footnote{\label{fermi} \url{https://fermi.gsfc.nasa.gov/ssc/data/access/lat/BackgroundModels.html}} and \texttt{iso\_P8R3\_SOURCE\_V3\_v1.txt}\footref{fermi}. In data processing the instrumental response function "P8R3\_SOURCE\_V3" was used. In likelihood analysis, the unbinned likelihood analysis\footnote{\url{https://fermi.gsfc.nasa.gov/ssc/data/analysis/scitools/likelihood_tutorial.html}} was performed using \texttt{GTLIKE} tool \citealt{cash1979parameter, mattox1996likelihood} that provides the significance of each source within the ROI including the source of interest in the form of test statistics. The Test Statistic is defined as TS = -2ln$\left(\frac{L_{max,0}}{L_{max,1}}\right)$, where $L_{max,0}$ and $L_{max,1}$ are the maximum likelihood value for a model without an additional source and the maximum likelihood value for a model with the additional source at a specified location, respectively.

We filtered out the sources with low TS i.e. below TS = 9. Using this criterion, a weekly binned lightcurve with TS$\> (\ge 9)$ was generated. In the process of lightcurve generation, parameters for sources positioned beyond 10$^\circ$ from the center of ROI were fixed, while for $\le 10^\circ$ were left unrestricted and were allowed to vary freely. In this study, We employed the \texttt{FERMIPY}\footnote{\url{https://fermipy.readthedocs.io/en/latest/}} to generate the light curve.
\begin{figure*}[ht!]
\plotone{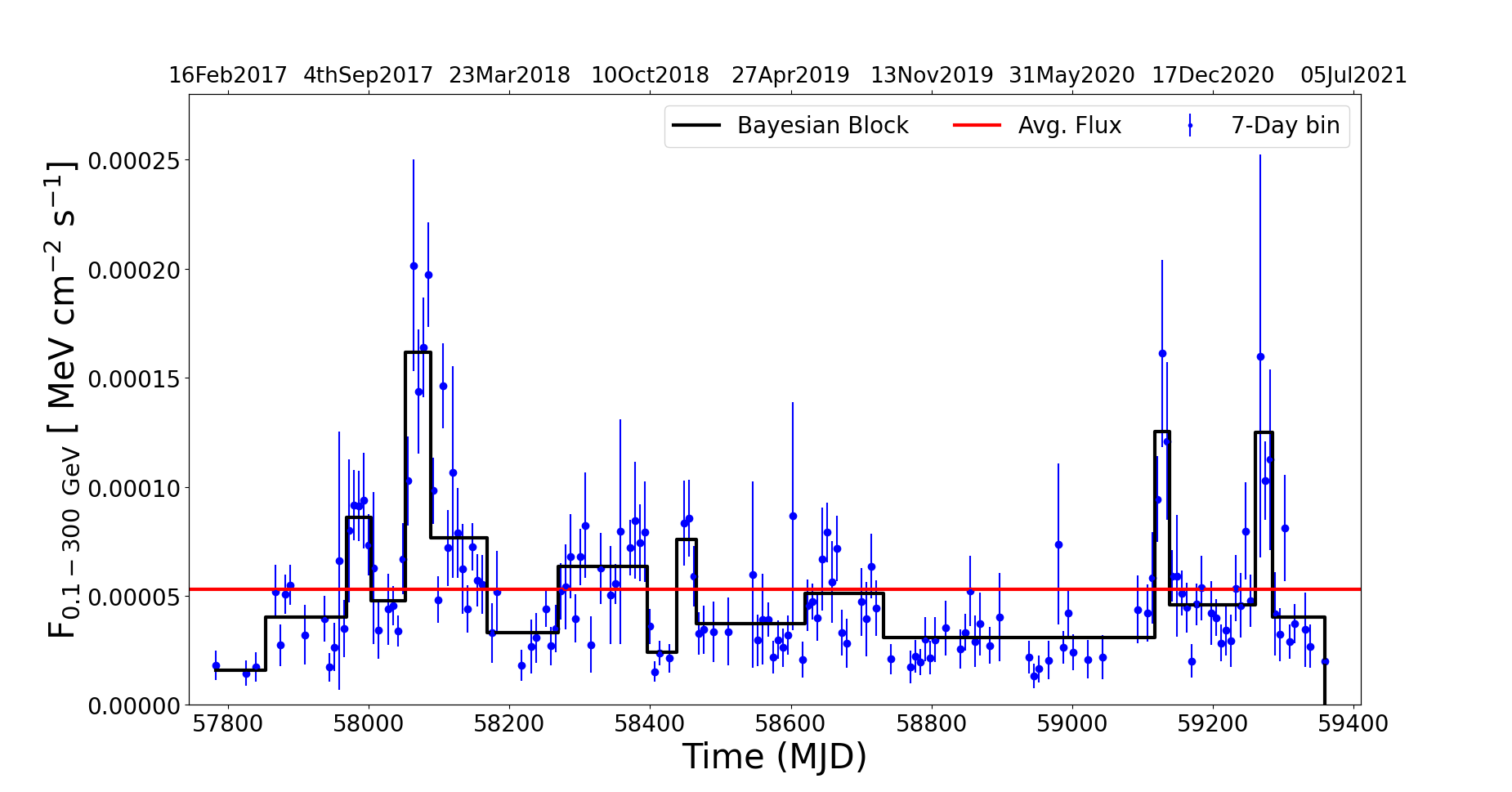}
\caption{Weekly binned Fermi-LAT light curve of PKS 2255-282 (MJD 57783.5 - 59358.5). Bayesian block representation of the light curve is shown by a black solid line.}
\label{fig:lightcurve}
\end{figure*}

\begin{figure*}
    \centering
    \includegraphics[width=1\linewidth]{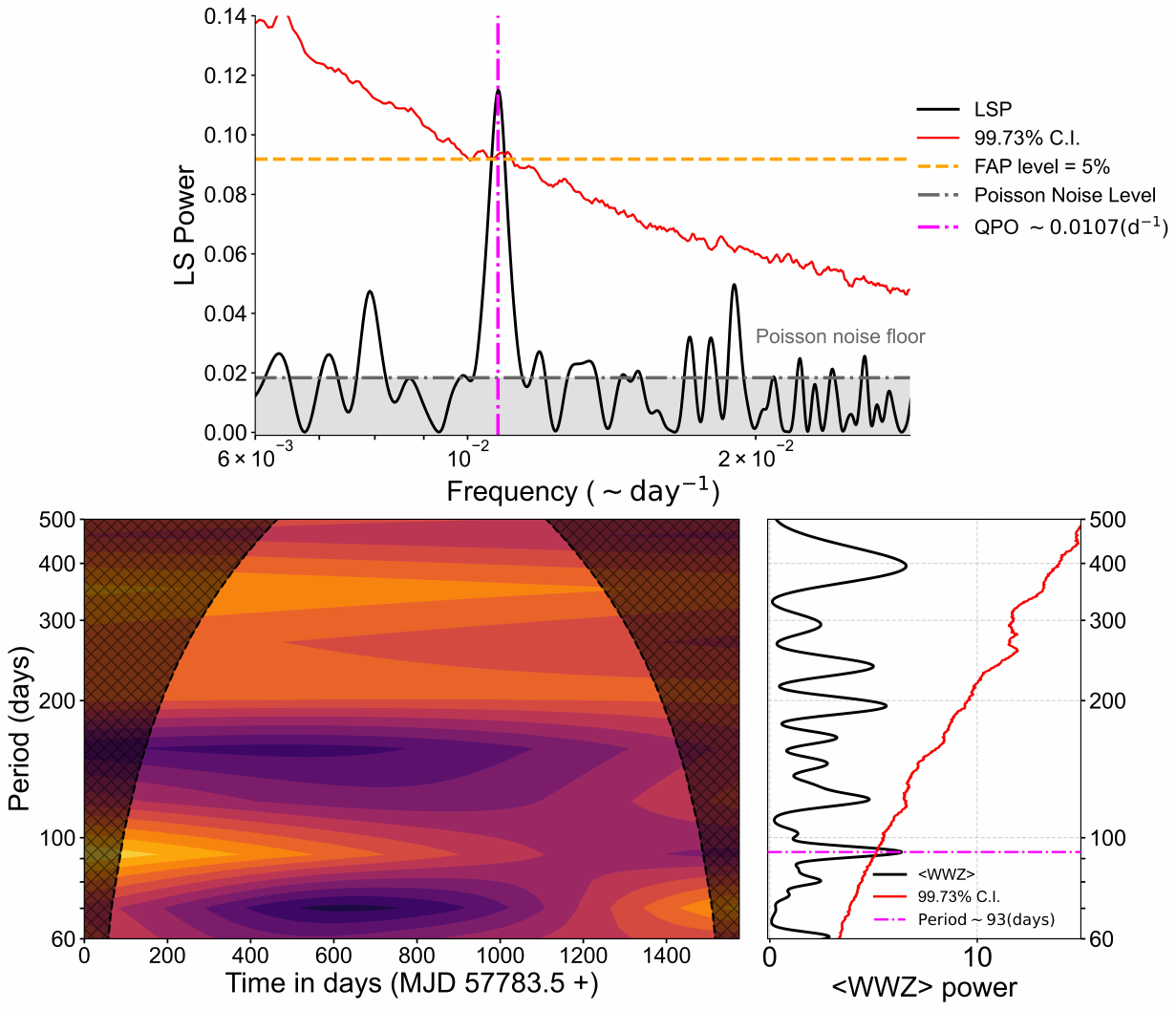}
    \caption{A transient QPO was examined with LSP and WWZ methods here. Top Panel: the LSP is depicted by the solid black line with a 99.73$\%$ significance line in red. Notably, the LSP exhibits a peak at $\sim$0.0107 $\text{d}^{-1}$ with a confidence level surpassing 3$\sigma$. In addition, a False Alarm Probability (FAP) has been computed and the value is 0.000406, and 95$\%$ confidence is visually represented with a horizontal dashed orange line, indicating 5$\%$ FAP level. Furthermore, The Poisson noise level is calculated to be approximately 0.0183, represented in a horizontal grey line. Bottom right: a WWZ map with the cone of influence (dark-shaded region) is depicted. Bottom Left: The solid black line represents the average wavelet power across time, while the red line denotes the 99.73$\%$ level of significance.}
    \label{fig:LSP_WWZ}
\end{figure*}

\section{\textbf{METHODOLOGIES}} \label{method}
The Figure \ref{fig:lightcurve} illustrates the weekly binned gamma-ray light curve of PKS 2255-282, along with the optimal Bayesian block representation.\par
We employed various methodologies and tests to search for periodicity in $\gamma$-ray lightcurve of the source. These included the \texttt{Lomb-Scargle periodogram (LSP)} and \texttt{Weighted Wavelet Z-transform (WWZ)}. Furthermore, we incorporated statistical models to perform time series modeling, namely the \texttt{Autoregressive Integrated Moving Average (ARIMA)} and \texttt{Seasonal Autoregressive Integrated Moving Average (SARIMA)}. To expand our exploration of periodicity, Gaussian Process modeling was introduced, using two distinct models: the \texttt{Stochastically Driven Damped Harmonic Oscillator (SHO)}. By applying these various approaches, we meticulously examined the $\gamma$-ray lightcurve of the source, and the results of our investigations are detailed in the subsequent sections.\par

\subsection{\textbf{Lomb-Scargle periodogram}}
The Lomb-Scargle periodogram (LSP) \citep{lomb1976least, scargle1982studies} is a widely used method to identify possible periodic patterns in time series data. This approach involves fitting a sinusoidal function to the data using a least square method. What makes this technique advantageous is its ability to handle unevenly sampled data effectively, minimize the impact of noise, and offer a precise measurement of the identified periodicity in the time series. In this study, we computed the LSP using the \texttt{LOMB-SCARGLE}\footnote{\url{https://docs.astropy.org/en/stable/timeseries/lombscargle.html}} package provided by \texttt{Astropy}. The power of the LSP is given by \citep{vanderplas2018understanding}:\par

\begin{equation}
\begin{split}
P_{LS}(f) = \frac{1}{2} \bigg[ &\frac{\left(\sum_{i=1}^{N} x_i \cos(2\pi f (t_i - \tau))\right)^2}{\sum_{i=1}^{N} \cos^2(2\pi f (t_i - \tau))} \\
&+ \frac{\left(\sum_{i=1}^N x_i \sin(2\pi f (t_i - \tau))\right)^2}{\sum_{i=1}^N \sin^2(2\pi f (t_i - \tau))} \bigg]
\end{split}
\end{equation}

where, $\tau$ is 
\begin{equation}
    \tau = \rm{tan^{-1}} \left(\frac{\sum_{i=1}^N sin\left(2\pi f (t_i - \tau) \right)}{2 (2\pi f) \sum_{i=1}^N cos\left(2\pi f (t_i - \tau) \right)} \right)
\end{equation}

In this study, we selected the minimum ($\rm{f_{min}}$) and maximum ($\rm{f_{max}}$) values for the temporal frequency as $1/T$ and $1/\rm{2\Delta T}$, respectively. Here, $T$ represents the total observation period, and $\Delta{T}$ is the time difference between two consecutive points.\par 
Furthermore, it is a common practice to assess the existence of periodic patterns using the \texttt{Generalized Lomb-Scargle periodogram\footnote{\url{https://pyastronomy.readthedocs.io/en/latest/pyTimingDoc/pyPeriodDoc/gls.html}} (GLSP)}. This method also accounts for measurement uncertainties during the analysis. The results of this analysis offer additional confirmation of periodicity, reinforcing our findings.\par

The LSP analysis reveals a peak of period 92.9$\pm$2.5 days with the significance 4.06$\sigma$. The error on the observed period is calculated by fitting a Gaussian to the dominant LSP peak, and half-width at half-maximum (HWHM) was used as an error in period value. The observed time scale of transient QPO is consistent with the WWZ finding within the error bar. In the top panel of  Figure \ref{fig:LSP_WWZ}, the red solid line represents the 3$\sigma$ significance level.

\subsection{\textbf{Weighted Wavelet Z-transform}}
The Weighted Wavelet Z-transform (WWZ) is a robust method used in astronomical studies to identify transient periodic patterns in irregularly sampled time series data. WWZ convolves the light curves with the time and frequency-dependent kernel and attempts to localize the periodicity feature in temporal and spectral space, known as the ``WWZ Map". In studying the evolution of QPO features over time, WWZ emerges as a powerful tool, enabling us to identify how these oscillations gradually develop, evolve, and eventually fade over time \citep{foster1996wavelets}.\par

In wavelet analysis, we used the abbreviated Morlet kernel, which has the following form:
\begin{equation}
    f[\omega (t - \tau)] = \exp[i \omega (t - \tau) - c \omega^2 (t - \tau)^2]
\end{equation}

and the corresponding WWZ map is given by,
\begin{equation}
    W[\omega, \tau: x(t)] = \omega^{1/2} \int x(t)f^* [\omega(t - \tau)] dt
\end{equation}
Here, $f^*$ is the complex conjugate of the wavelet kernel $f$; $\omega$ and $\tau$ are the frequency and the time-shift, respectively. This kernel acts as a windowed DFT, where the size of the window is determined by both the parameters $\omega$ and a constant \textit{c}. The resulting WWZ map offers a notable advantage, it not only identifies dominant periodicities but also provides insights into their duration over time. Additionally, we also incorporated the Cone of Influence (COI), see Figure \ref{fig:LSP_WWZ}. It is necessary to consider the impact of edge effects in wavelet analysis. This region in time-frequency space signifies where edge effects become significant, making it harder to discern specific frequencies due to the decreasing number of data points in the wavelet. In practical application, we often deal with finite time series. As the wavelet approaches the edge, this number decreases, because detecting particular frequencies depends on the number of data points in the sampled frequency regime, impacting the reliability of the detected frequency or period near the border. In this work, we use a grey-shaded region to indicate the cone of influence in the WWZ plots. \par


\begin{figure}
    \centering
    \includegraphics[width=0.99\linewidth]{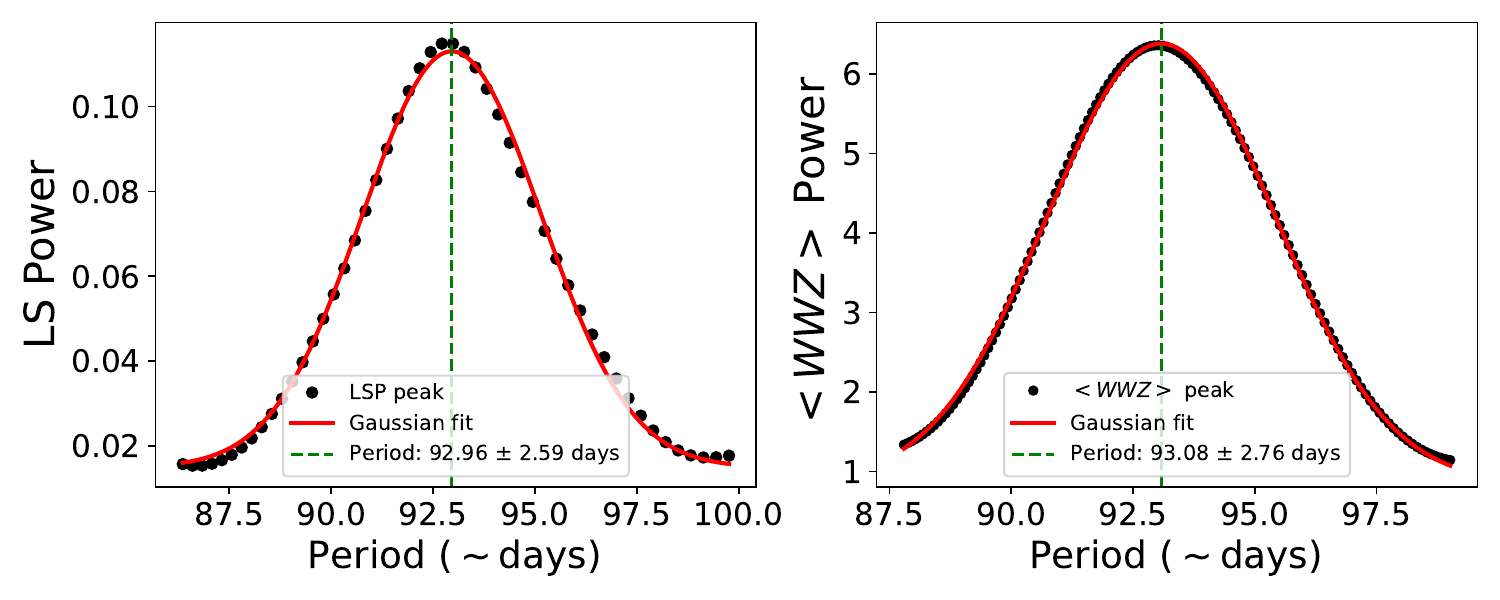}
    \caption{The uncertainty in the observed period from LSP and WWZ analysis was estimated by fitting the dominant peaks with a Gaussian function. Left Panel: The LSP peak with a period of 92$\pm$2.5 days. Right Panel: The $<WWZ>$ power plot reveals a peak with a period of 92$\pm$2.7 days. A green vertical line represents the peak position of best fit.}
    \label{fig:period_error}
\end{figure}

\begin{figure}
    \centering
    \includegraphics[width=0.9\linewidth]{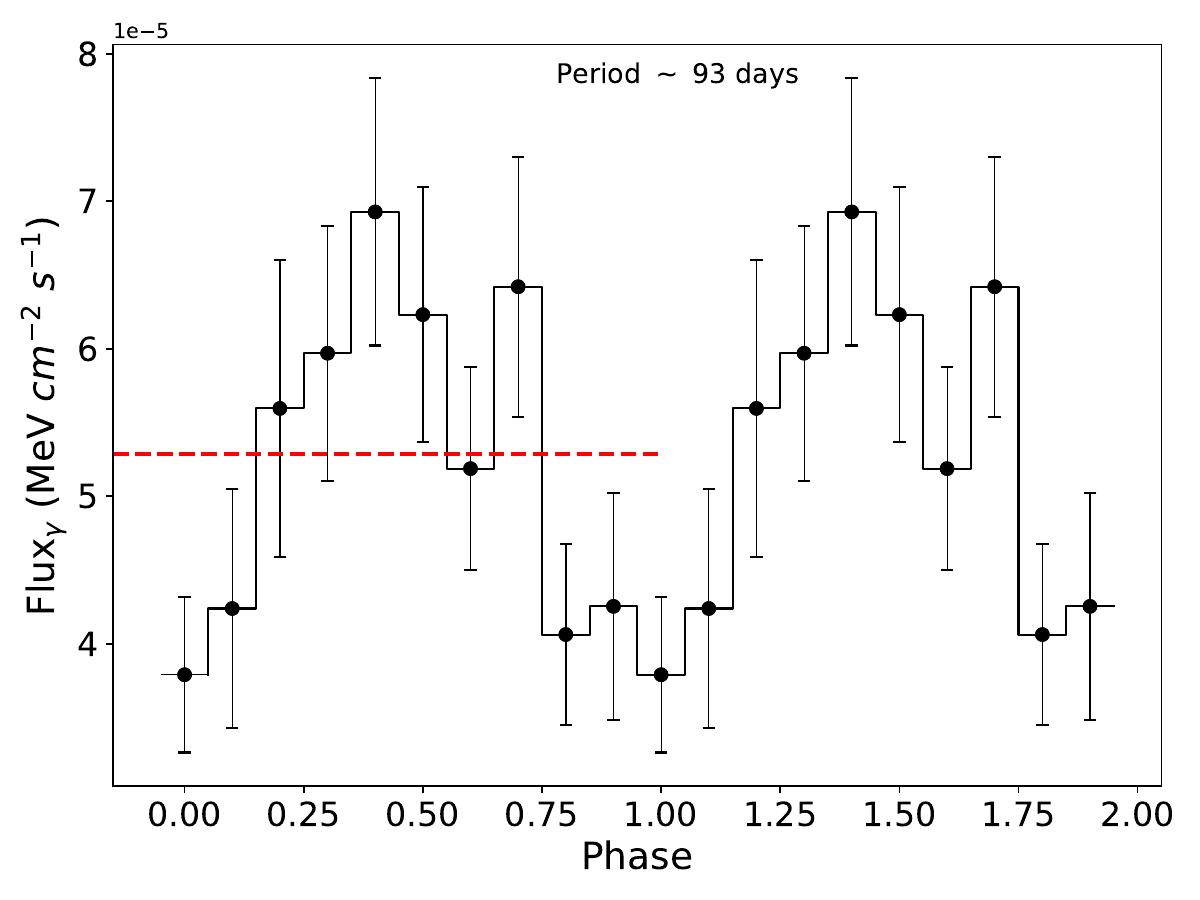}
    \caption{The folded Fermi-LAT light curve of PKS 2255-282 in the time domain from MJD 57783.5 - 59358.5 above 100 MeV with a period of 93 days. The dashed red line represents the mean value. We present two-period cycles here for better clarity.}
    \label{fig:folded_lc}
\end{figure}

For this study, we used publicly available Python code\footnote{\url{https://github.com/eaydin/WWZ}}, \cite{m_emre_aydin_2017_375648}, to generate the WWZ map. From the analysis, we observed a power concentration around $\sim$0.0107 ($d^{-1}$), corresponding to 93$\pm$2.7 days, in the WWZ map. The uncertainty in the period was calculated as described in section 3.1. The observed peak in the time-averaged WWZ has demonstrated a significance level of 3.96$\sigma$. Notably, as observed, the Fourier-based analysis revealed a 93-day QPO signature in the light curve.

\subsection{\textbf{Significance test}}
The presence of red noise in the AGN light curve motivated us to assess the significance of the periodic features using the Monte-Carlo method developed by \citep{emmanoulopoulos2013generating}. The method involves modeling the Power Spectral Density (PSD) generated through a simple power-law model. We conducted a Monte Carlo simulation by generating 1$\times 10^5$ synthetic light curves using DELightcurveSimulation\footnote{\url{https://github.com/samconnolly/DELightcurveSimulation}}, each mimicking the underlying properties of the original data.

Additionally, the Baluev approximation method is used to estimate the False Alarm Probability (FAP), which is defined as;

\begin{equation}
    FAP(P_n)=1 - (1 - Prob(P > P_n))^M
\end{equation}

where the FAP represents the probability that at least one out of M independent power values within a specified frequency band of a white noise periodogram will be greater than or equal to the power threshold ($\rm{P_n}$). To stabilize the M independent trials, which can be defined as $M = \frac{\Delta f}{\delta f}$, where $\Delta f = f_{nyq} - f_{min}$, is frequency range and $\delta f = \frac{1}{T}$, is frequency resolution, in which T represents the total period of observation.
Besides the significance test, the Poisson noise level was calculated, arising from the statistical noise due to the uncertainty associated with measurement, $x_i$. The expression for the same is given by 

\begin{equation}
    P_{noise} = \frac{2 T \bar{\Delta x^2}}{N^2 \bar{x}^2}
\end{equation}
where $T$ denotes the total light curve duration, $N$ represents the total number of measurements $x_i$. $\bar{x}$ and $\Delta x$ represent the mean flux and uncertainty in the measurements, respectively. The calculated Poisson noise level is approximately 0.0183.\par

In order to further confirm the presence of the QPO signature, we run a folding search of the source light curve. This method yielded a modulation with a
period of 93 days. The presence of this modulation lends further credence towards our Fourier-based QPO finding results (see Figure \ref{fig:folded_lc}).

\begin{figure*}
    \centering
    \includegraphics[width=.5\linewidth]{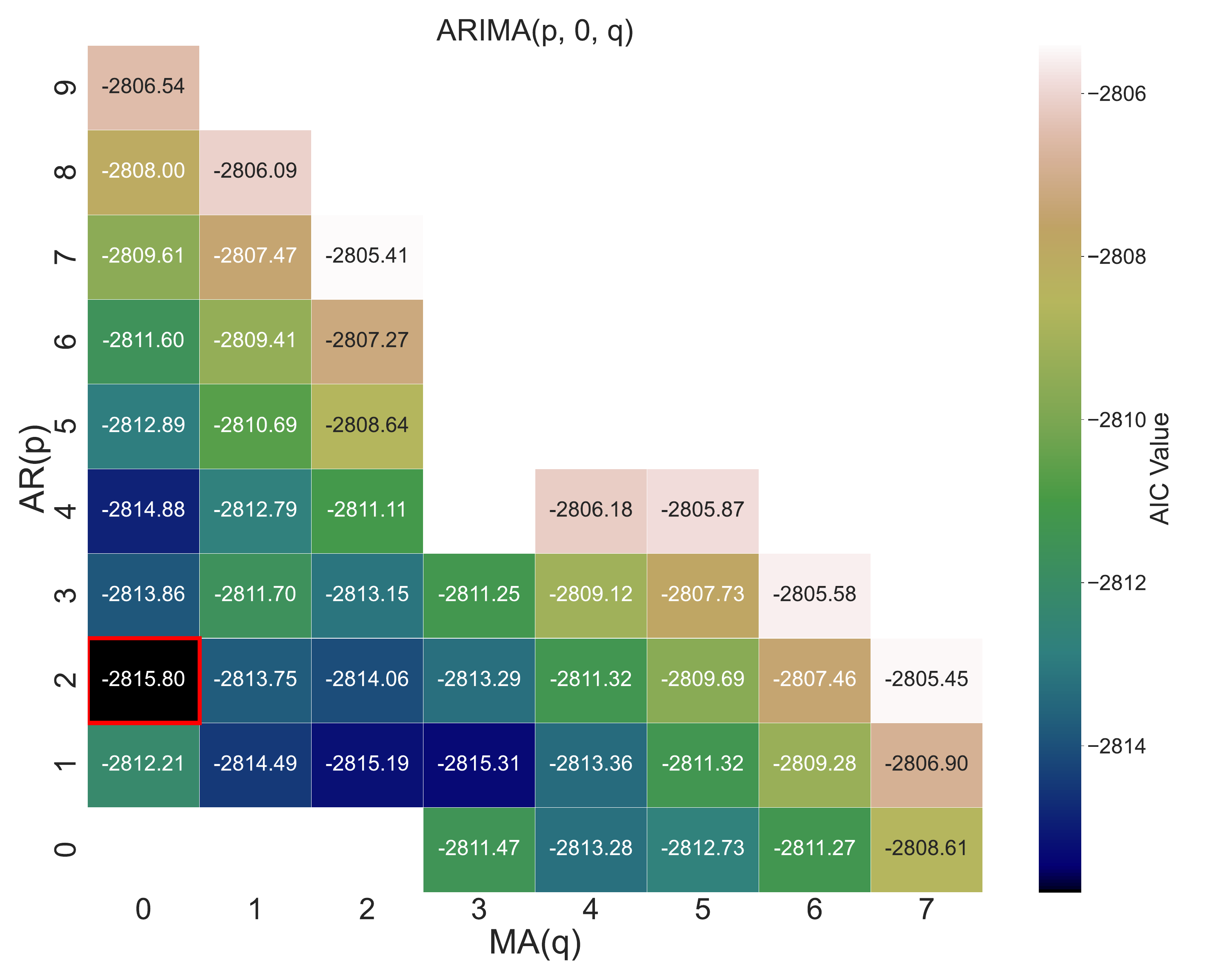}\hfill
    \includegraphics[width=.5\linewidth]
    {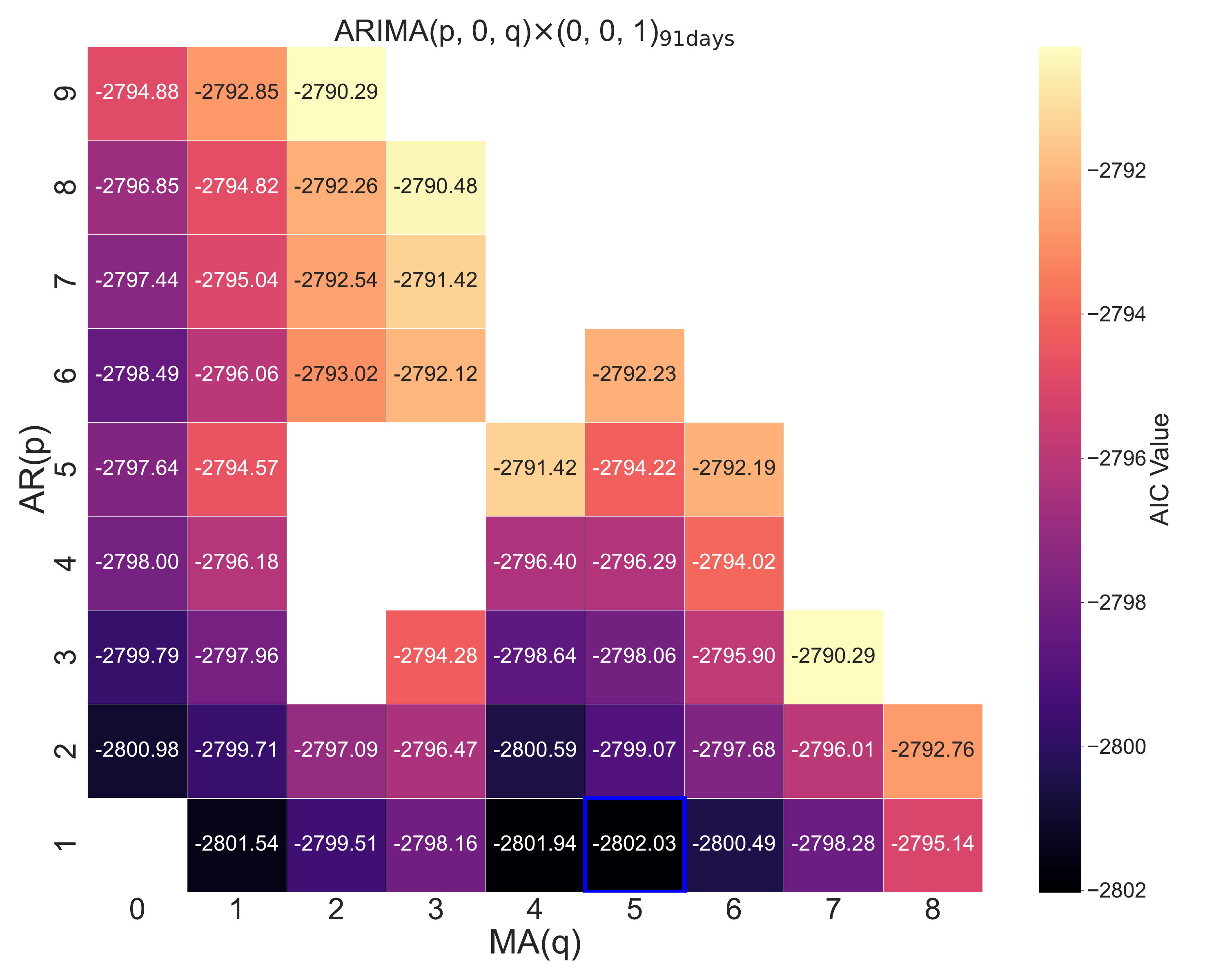}\vfill
    \includegraphics[width=.65\linewidth]{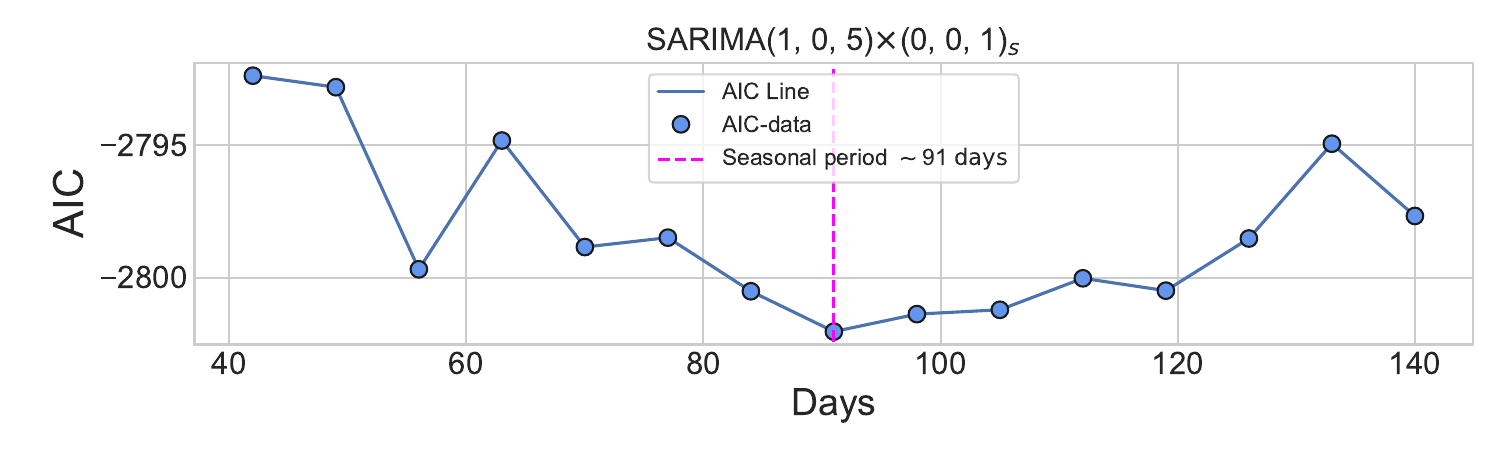}
    \caption{In the left panel, we present the AIC distribution map of ARIMA models for the source PKS 2255-282. The optimal model, ARIMA(2, 0, 0), is highlighted in red, with an AIC of -2815.80. On the right, the AIC map for SARIMA models shows that SARIMA(1, 0, 5)×(0, 0, 1)$_{s=91}$ is identified as the best model, with an AIC value of -2802.03. In the bottom panel, the AIC values for this SARIMA model were explored for different periods, and the model achieved its global minimum AIC at s=91 days. The blue dots represent AIC values at various seasonal positions.}
    \label{fig:arima_sarima}
\end{figure*}

\begin{figure}
    \centering
    \includegraphics[width=0.9\linewidth]{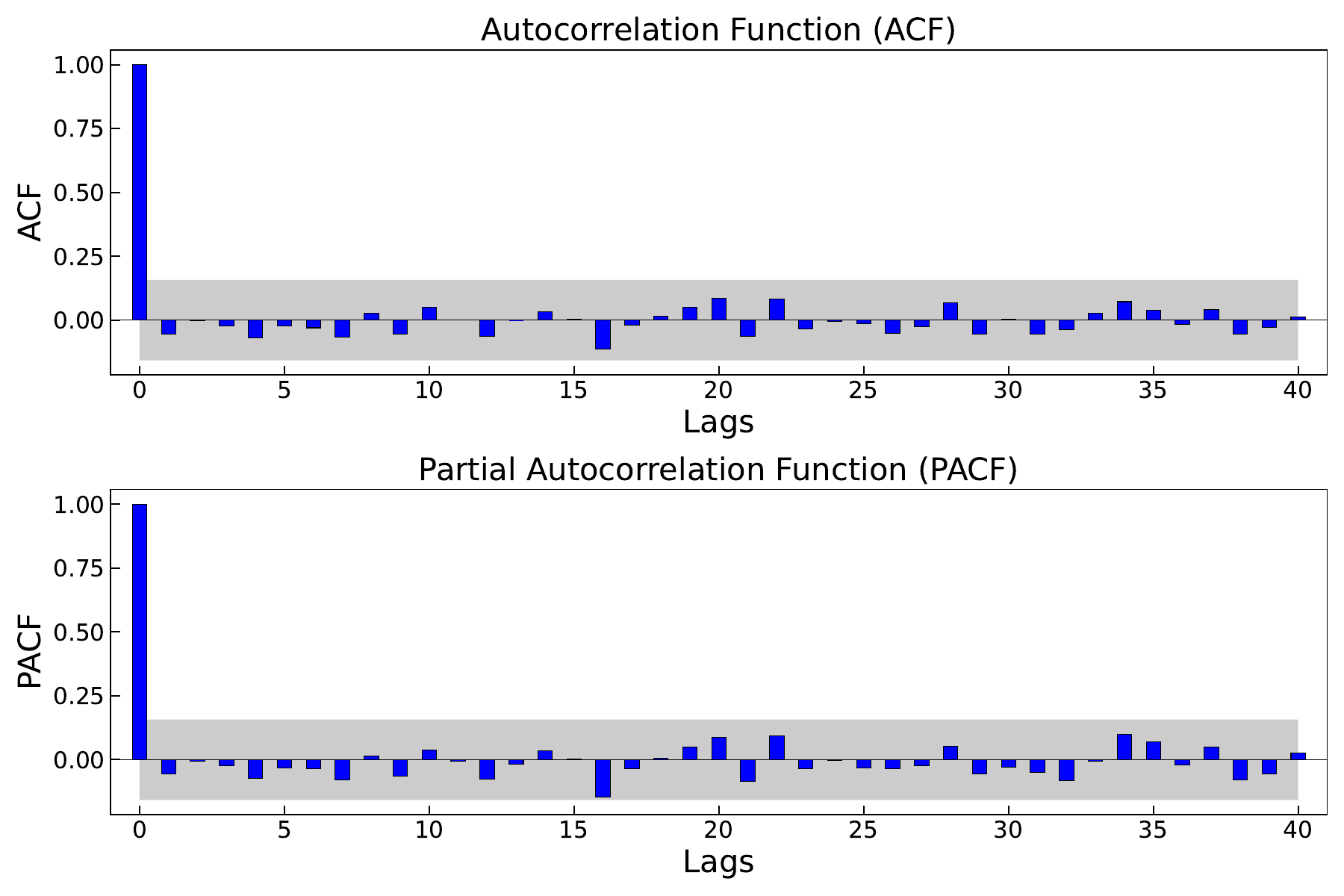}\vfill
    \includegraphics[width=0.9\linewidth]{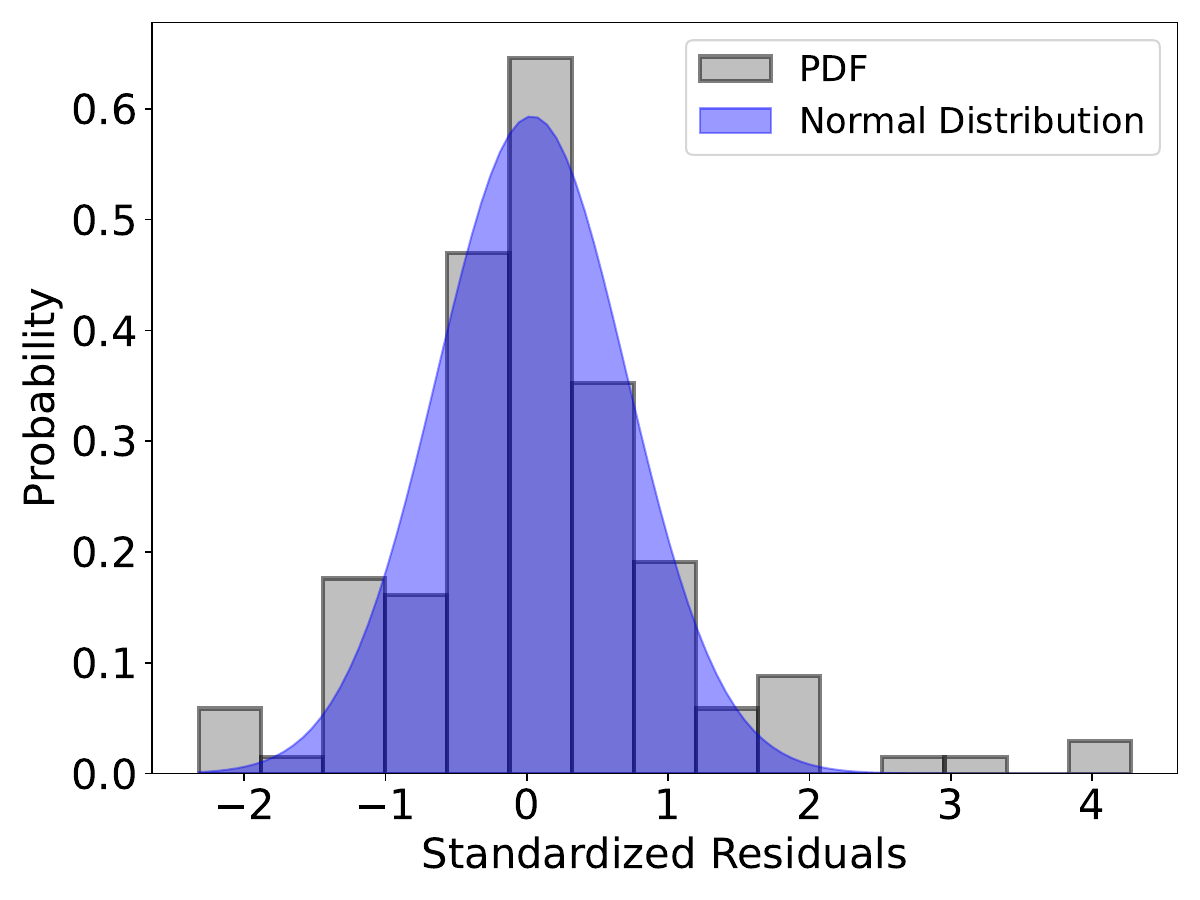}
    \caption{The Autocorrelation function and Partial Autocorrelation function of standardized residuals are depicted in the top two panels and the bottom panel represents a histogram of standardized residuals fitted with normal distribution.}
    \label{fig:std_res}
\end{figure}

\subsection{\textbf{Time series modeling}}
The study of astrophysical objects through timing analysis enables astronomers to investigate their evolutionary patterns and dynamic properties. By breaking down time series into various components associated with different physical processes, temporal analysis offers a means of understanding the underlying mechanisms that give rise to time-varying signals. Through a range of methods and techniques, astronomers can explore the distinctive features of these components and estimate parameters that allow for the recovery of deterministic properties.\

We used the autoregressive integrated moving average (ARIMA) models\footnote{\url{https://www.statsmodels.org/dev/generated/statsmodels.tsa.arima.model.ARIMA.html}} to model the light curve. This model is commonly employed in various fields. The ARIMA model comprises three main components: autoregressive (AR), integrated (I), and moving average (MA) processes. Considering $\textit{F}(t_i)$ to be the emission at time $t_i$ and $\epsilon (t_i)$ represents the fluctuations. The analytical form of the AR time series is $ \textit{F}(t_i) = \sum_{j=1}^p \theta_j \textit{F}(t_{i - j}) + \epsilon (t_i)$, where p is the order of the AR process. The AR process determines the coefficient of dependence between current and past emissions and $\theta_j$'s are the AR coefficients. The integrated (I) process reduces trends, and the MA process quantify the coefficient of current emission's dependence on the system's recent random shocks \citep[]{scargle1981studies,feigelson2018autoregressive}, an expression of Moving Average (MA) model for the time series is $\textit{F}(t_i) = \sum_{j=1}^q \phi_j \epsilon (t_{i - j}) + \epsilon (t_i)$, where q is the order of the MA and $\phi_j$'s are the MA coefficients.\par
The analytical representation of the ARIMA(p, d, q) model is given as :
\begin{equation}
    \Delta^d F(t_i) = \sum_{j=1}^{p} \theta_j \Delta^d F(t_{i-j}) + \sum_{j=1}^{q} \phi_j \epsilon(t_{i-j}) + \epsilon(t_i),
\end{equation} or
\begin{equation}
    \left( 1 - \sum_{j=1}^{p} \theta_j L^j \right) \Delta^d F(t_i) = \left (1 - \sum_{j=1}^{q} \phi_j L^j \right)\epsilon(t_i),
    \label{eq:ARIMA}
\end{equation}
where p, d, and q are the AR order, order of differencing, and MA order respectively. Further extension of the ARIMA model to identify physical periodicity in the light curve by including seasonal features, resulting in Seasonal(S) ARIMA(p, d, q) models or SARIMA$(p, d, q)\times(P, D, Q)_s$ models\footnote{\url{https://www.statsmodels.org/dev/generated/statsmodels.tsa.statespace.sarimax.SARIMAX.html}} \citep[]{adhikari2013introductory, permanasari2013sarima, sarkar2021multiwaveband, chen202231}. The representation of SARIMA$(p, d, q)\times(P, D, Q)_s$ model is defined as
\begin{equation} \label{eq:SARIMA}
\begin{split}
\left(1 - \sum_{j=1}^{p} \theta_j L^j \right)\left(1 - \sum_{j=1}^{P} \Theta_j L^{sj}\right)\Delta^d \Delta_{s}^{D}F(t_i) \\
= \left(1 + \sum_{j=1}^{p}\theta_j L^j \right)\left(1 + \sum_{j=1}^{Q}\Phi_j L^{sj}\right)\epsilon(t_i) + A(t),
\end{split}
\end{equation}

where P, D, Q, and s are the seasonal AR order, order of differencing, MA order, and seasonal parameter respectively. To determine the best model for our time series, we employed both the autoregressive integrated moving average (ARIMA) and seasonal ARIMA (SARIMA) models and compared their goodness of fit. We used the Akaike information criterion (AIC) to assess the relative quality of each model. The AIC is calculated as -2$ln L$ + 2$k$, where $L$ is the likelihood function and k is the number of free parameters in the model. By comparing the AIC values of different models, we were able to identify the best-performing model. The model with the lowest AIC value was considered the most suitable for our time series analysis. Therefore, we build the parameter space to search for the model with the smallest AIC,

\begin{equation}
  \psi =
    \begin{cases}
      p, q    \in [0,9]\\
      P, Q    \in [0,6]\\
      d, D    \in [0,1]\\
         s     \in [0, 16] \times 7 days\\
    \end{cases}       
\end{equation}

Figure \ref{fig:arima_sarima} shows the fitting results of ARIMA and SARIMA modeling.\par

In the stochastic modeling of the light curve with ARIMA and SARIMA models, ARIMA(2, 0, 0) is a non-seasonal model with AIC -2815.8, and SARIMA(1, 0, 5)$\times$(0, 0, 1)$_{91}$ is a seasonal ARIMA model for the light curve with AIC -2802.3. The bottom panel of Figure \ref{fig:arima_sarima} presents the AIC values considering different periods, and we found that the best AIC occurs at the 91 ± 3.5 day position. The uncertainty in the seasonal period is estimated to be half of the light-curve time bin (3.5 days). To assess the goodness of the model, we observed that all spikes in the Autocorrelation function (ACF) and Partial Autocorrelation function (PACF) fall within the 95$\%$ confidence intervals of white noise. The observed mean value of the distribution is 0.014, see Figure \ref{fig:std_res}. The normality test using the Kolmogorov-Smirnov (KS) test\footnote{\url{https://docs.scipy.org/doc/scipy/reference/generated/scipy.stats.kstest.html}} yielded a p-value of 0.184, indicating that we cannot reject the null hypothesis that the sample is normally distributed. The findings from this modeling are consistent with LSP and WWZ results, further strengthening the reported QPO.

\begin{figure*}
    \centering
    \includegraphics[width=1\linewidth]{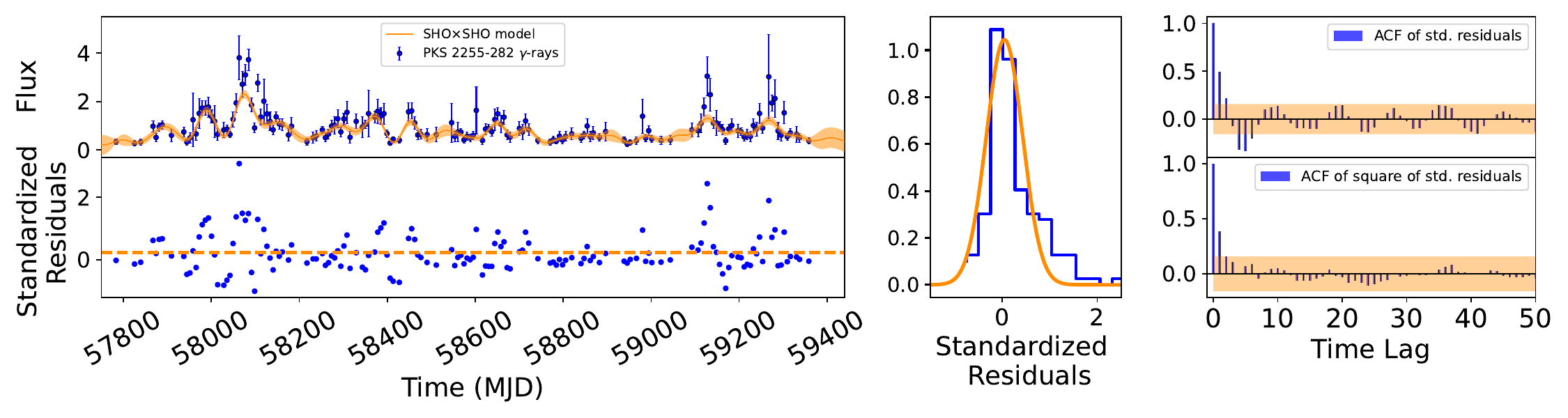}
    \caption{The \texttt{celerite} modeling was performed utilizing a 7-day binned $\gamma$-ray light curve of PKS 2255-282, spanning from MJD 57783.5 to 60114.5, using the SHO$\times$ 2 model. The top left panel shows the flux points and their uncertainties in blue, along with the best-fit profile from the \textit{celerite} modeling in orange, including the 1$\sigma$ confidence interval. The bottom left panel presents the standardized residuals as blue dots, with the horizontal orange dotted line indicating the mean of the standardized residuals. The middle panel features a histogram of the scaled standardized residuals in blue, accompanied by a solid orange line representing the expected scaled normal distribution. The right top and right bottom panels display the auto-correlation functions (ACFs) of the standardized residuals and the squared standardized residuals, respectively, with 95$\%$ confidence intervals of the white noise.}
    \label{fig:SHO2_celerite}
\end{figure*}


\begin{figure*}
    \centering
    \includegraphics[width=0.5\linewidth]{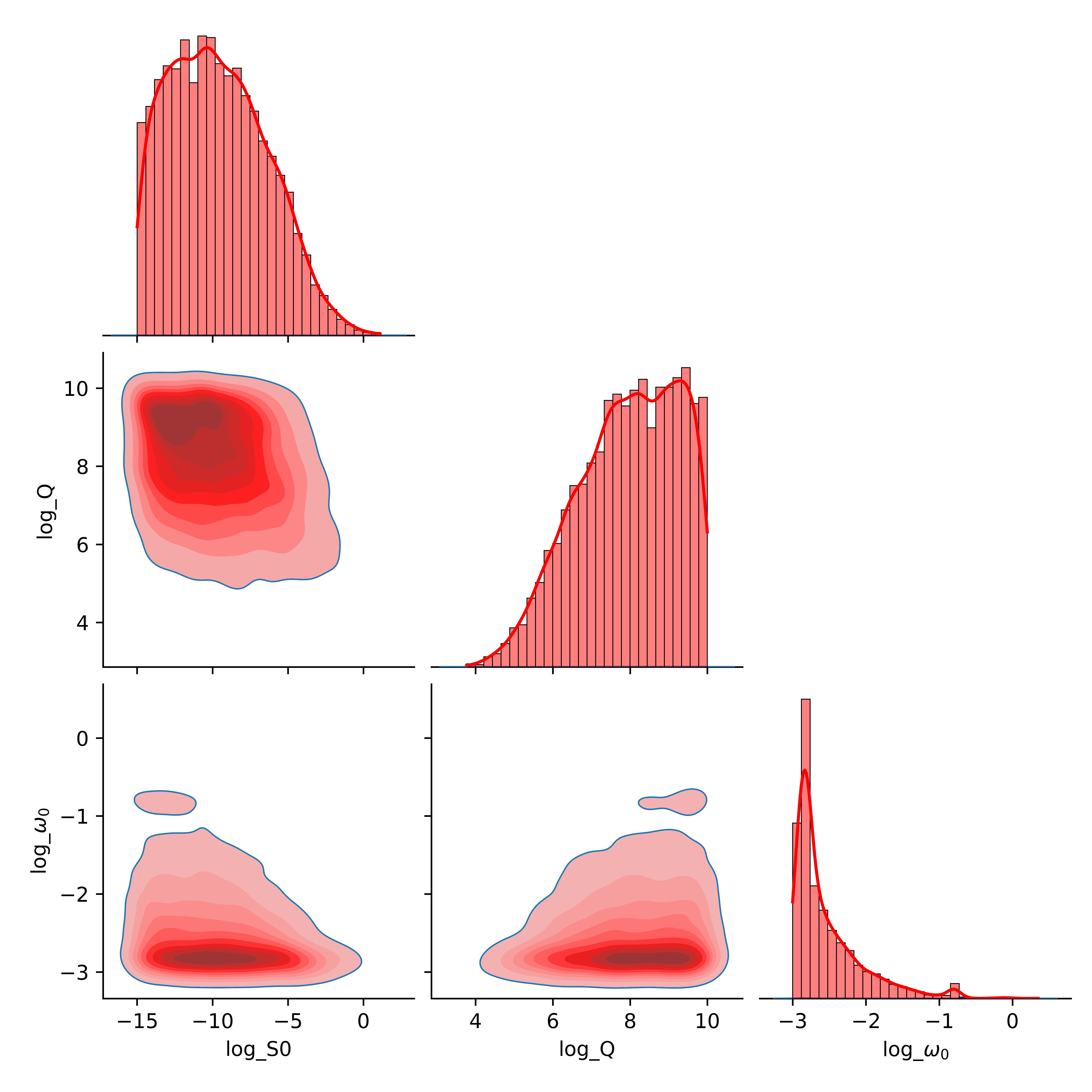}\hfill
    \includegraphics[width=0.5\linewidth]{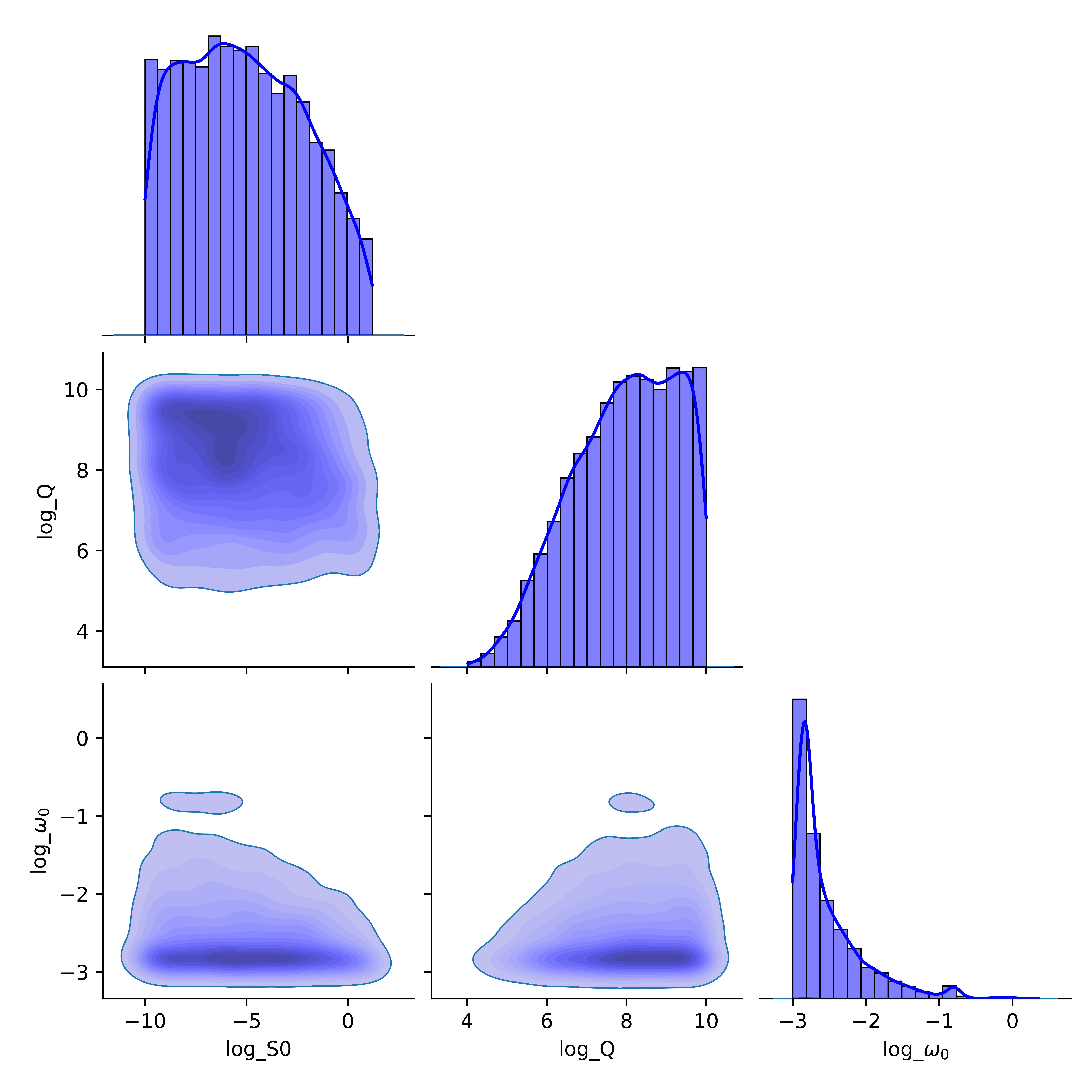}
    \caption{The figure displays the posterior probability densities of parameters obtained from the Celerite modeling with the SHO$\times$2 model. The left panel (red color) corresponds to the SHO$_1$ model, while the right panel corresponds to the SHO$_2$ model.}
    \label{fig:posterior}
\end{figure*}

\begin{figure}
    \centering
    \includegraphics[width=1\linewidth]{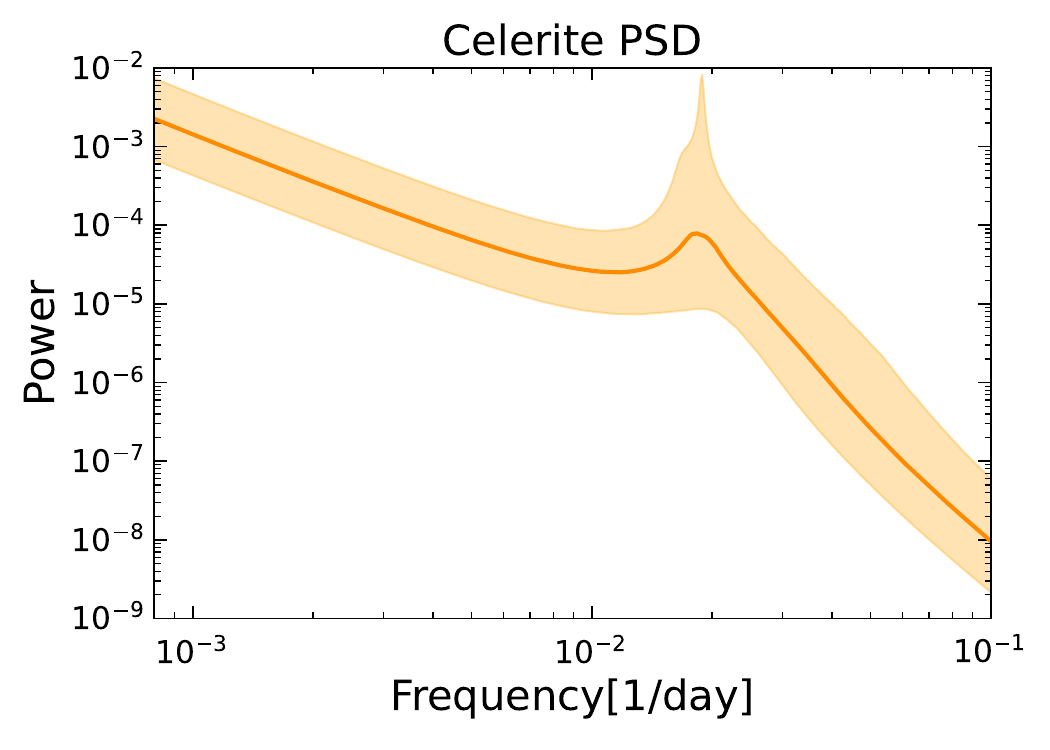}
    \caption{A PSD constructed from the celerite modeling results with SHO$\times$2 model for the 7-day binned $\gamma$-ray light curve of PKS 2255-282 during MJD 57783.5 to 60114.5. The figure displays a plausible QPO feature with the period of $\sim$93 days with the 68$\%$ confidence band.}
    \label{fig:celerite_psd}
\end{figure}
\subsection{\textbf{Gaussian Process modelling}}
In addition to approaches based on frequency domain analysis  such as LSP and WWZ, and time series modeling with statistical models for analyzing astronomical variability, an alternative method involves employing Gaussian process modeling of light curves in the temporal domain. In this investigation, a Stochastically Driven Damped Harmonic Oscillator (SHO) from \textit{celerite}\footnote{\url{https://celerite.readthedocs.io/en/stable/}} \citep{foreman2017fast} is used, as it can describe the variability characteristics influenced by noisy physical processes, which grows strongly at a characteristic timescale but is also damped due to dissipation in the system. The differential equation for this system is  

\begin{equation}
    \left[ \frac{d^2}{dt^2} + \frac{\omega_0}{Q}\frac{d}{dt} + \omega_{0}^{2} \right]y(t) = \epsilon(t)
\end{equation}

where, $\omega_0$, Q, and $\epsilon$(t)  are the frequency of the undamped oscillator, the quality factor of the oscillator, and a stochastic driving force, respectively. If the $\epsilon$(t) is white noise, the PSD of this process is given as

\begin{equation}
    S(\omega) = \sqrt{\frac{2}{\pi}} \frac{S_0 \omega_{0}^4}{\left( \omega^2 - \omega_{0}^2\right)^2 + \frac{\omega^2 \omega_{0}^2}{Q^2}}
\end{equation}
here $S_0$ is proportional to the power at $\omega = \omega_0$, $S(\omega_0) = \sqrt{\frac{2}{\pi}}S_0 Q^2$.\par

To model light curves, an approach involves constructing a model using a combination of a set number of SHO terms and choosing log-uniform parameters space on each of four parameters as listed in Table \ref{tab:prior} . The fitting process employs the Markov Chain Monte Carlo (MCMC) algorithm provided by \textit{emcee}\footnote{\url{https://emcee.readthedocs.io/en/stable/}} \cite{emcee}. By sampling from MCMC, we can determine the model parameters' values and uncertainties. In the analysis, 2$\times 10^4$ samples are generated, in which 5$\times 10^3$ initial samples are discarded as burn-in, and the subsequent 1.5$\times 10^4$ samples are utilized for the final MCMC output. The obtained results of stochastic modeling are listed in Table \ref{tab:posterior}.

\begin{deluxetable}{cc}
\setlength{\extrarowheight}{6pt}
\setlength{\tabcolsep}{40pt}
\tablecaption{Parameters and Priors for celerite modeling.
\label{tab:prior}}
\tablehead{
\colhead{Parameter} & \colhead{Prior}
}
\startdata
$ln(S_0)$          & \textit{u}(-15,15) \\
$ln(Q)$            & \textit{u}(-15,15) \\
$ln(\omega_1)$   & \textit{u}(-15,15) \\
$ln(\omega_2)$   & \textit{u}(-10,10) \\
\enddata
\end{deluxetable}

\begin{deluxetable*}{cccc}
\setlength{\extrarowheight}{8pt}
\setlength{\tabcolsep}{30pt}
\tablecaption{Posterior parameter of celerite model.
\label{tab:posterior}}
\tablehead{
\colhead{Model} & \colhead{$\ln S_0$} & \colhead{$\ln Q$} & \colhead{$\ln \omega_0$}\\
(1) & (2) & (3) & (4)
}
\startdata
${   }$ & $-9.85_{-3.32}^{+3.72}$ & $8.09_{-1.55}^{+1.29}$ & $-2.72_{-0.16}^{+0.60}$ \\
$\rm{SHO}\times2 $ & $ $ & $ $ & $ $ \\
${   }$ & $-5.31_{-3.16}^{+3.57}$ & $8.09_{-1.55}^{+1.30}$ & $-2.72_{-0.16}^{+0.60}$ \\
\enddata

\tablecomments{The best-fitting parameters of the SHO$\times$2 model are shown here. (1) Model, (2)-(4) posterior parameters of the SHO$\times$2 model.}
\end{deluxetable*}


\begin{deluxetable*}{cccccc}
\setlength{\extrarowheight}{6pt}
\setlength{\tabcolsep}{6pt}
\tablecaption{The QPO period in days.
\label{tab:results}}
\tablehead{
\colhead{4FGL Name} & \colhead{Association Name} & \colhead{LSP} & \colhead{WWZ} & \colhead{SARIMA} & \colhead{Gaussian Modeling}\\
    &   &   &   & (1,0,5)$\times$(0,0,1,91) & SHO$\times$SHO \\
(1) & (2) & (3) & (4) & (5) & (6)
}
\startdata
4FGL J2258.1-2759 & PKS 2255-282 & 93$\pm$2.59 ($\approx$4.06$\sigma$) & 93$\pm$2.76 ($\approx$3.96$\sigma$) & 91$\pm$3.5 & 95$_{-43.05}^{+16.55}$ 
\enddata

\tablecomments{The observed QPO period(in days) from different methodologies. (1) Source name in the 4$^{th}$ Fermi-LAT catalog, (2) Associated name, (3) the observed QPO period in LSP analysis, (4) the observed QPO period in WWZ analysis, (5) The observed seasonal component in SARIMA modeling, (6) The observed QPO period in Gaussian modeling.}
\end{deluxetable*}


\section{\textbf{RESULT}} \label{results}
\label{result}

In this study, we analyzed the weekly-binned $\gamma$-ray light curve of the blazar PKS 2255-282 using various methods to detect potential periodic signals. The Lomb-Scargle Periodogram (LSP) analysis revealed a prominent peak at a frequency of approximately 0.0107 $d^{-1}$ (92.9$\pm$2.6 days). To assess the statistical significance of the observed peak, we conducted a Monte Carlo simulation, confirming its significance of 4.06$\sigma$ (Figure \ref{fig:LSP_WWZ}). Additionally, the peak exceeds a 5$\%$ false alarm probability (FAP) threshold, obtained Poisson noise level is 0.0183.\par

Another approach based on Fourier analysis is the Weighted Wavelet Z-Transform (WWZ), as shown in the bottom panels of Figure \ref{fig:LSP_WWZ}. The WWZ map displays a clear power concentration around the frequency of 0.0107 $d^{-1}$ (93$\pm$2.7). This bright region suggests a possible transient QPO with a period of 93 days. The time-averaged WWZ, plotted in the bottom right panel of Figure \ref{fig:LSP_WWZ}, confirms this finding. A similar Monte Carlo simulation was performed to determine the significance of the average WWZ peak, confirming the 93-day QPO with a significance level of 3.96$\sigma$.\par

Figure \ref{fig:arima_sarima} presents the results of fitting both ARIMA and SARIMA models to the light curve. The distribution of the ARIMA and SARIMA model is depicted in Figure \ref{fig:arima_sarima} and the optimal model was chosen based on the AIC value. The best non-seasonal model is ARIMA(2,0,0), which achieved the lowest AIC value of -2815.8. The best seasonal ARIMA model, SARIMA(1,0,5) × (0,0,1)$_{91}$, has an AIC value of -2802.03. To assess the goodness of fit, we performed a Kolmogorov-Smirnov (KS) test on the standardized residuals of the best-fitting model, obtaining a p-value of 0.184. This indicates that the residuals follow a normal distribution.

In Figure \ref{fig:arima_sarima}, the bottom panel, the AIC values at different periods are presented, revealing the optimal AIC at the position corresponding to a period of 91 days with an uncertainty of 3.5 days. The uncertainty associated with the seasonal value was estimated from half of the light curve time bin.\par

In Gaussian process modeling, the modeled light curve, the posterior parameters, and the observed power spectral density are shown in Figures \ref{fig:SHO2_celerite}, \ref{fig:posterior}, \ref{fig:celerite_psd}, respectively. The detected oscillatory feature has a period of 97 days, with results summarized in Table 2. To assess the normality of the residuals, we conducted the Kolmogorov-Smirnov test. The resulting p-value of 0.0030 suggests rejecting the null hypothesis of normality. This conclusion is further supported by the ACF plot, indicating possible nonlinear behaviors captured by the stochastic model within the time series.

All methods used in this study suggest the presence of a possible transient quasi-periodic oscillation (QPO) with a period of $\sim$ 93 days in the gamma-ray light curve within the MJD 57783 - 59358 range, summarized in Table \ref{tab:results}.

\section{\textbf{DISCUSSION AND CONCLUSIONS}} \label{discussion}
In this study, we report the detection of 93 days of periodicity in the gamma-ray light curve of PKS 2255-282 using three different approaches. What's particularly promising is that the 93-day quasi-oscillation is consistently detected in all three approaches: Fourier-based analysis with LSP and WWZ, stochastic autoregressive variability analysis using SARIMA models, and time domain analysis using Gaussian process models. The significance of the observed peak surpasses a 3$\sigma$ threshold. To further understand the periodic nature of $\gamma$-ray emissions, we applied stochastic autoregressive models with seasonal components and found a seasonal component value of 91$\pm$3.5 days with the global minimum AIC value among all models. This suggests the presence of a strictly periodic component in the light curve, in addition to any QPO features indicated by the autoregressive component. Additionally, we utilized Gaussian process modeling with the SHO$\times$2 model, and the results of the modeling also revealed a QPO feature with a period of  $\sim$97 days. This finding aligns with the results obtained from the other two methodologies. The consistency of the detected 93-day QPO in the $\gamma$-ray light curve of the source across all employed approaches is particularly promising.\par

The observed periodic modulations in light curves have been attributed to various physical processes. One possible explanation is a binary supermassive black hole (SMBH) system. In this model, the secondary black hole orbits around the primary black hole and passes through its accretion disc. This interaction could give rise to quasi-periodic features in the emitted radiation. A specific case reported a periodic modulation with a 12-year timescale in OJ 287 \citep{valtonen2008massive, villforth2010variability, sandrinelli2016quasi}, which was attributed to this binary SMBH scenario. However, the timescale observed in our work is much shorter, on the order of months, making it unlikely that this particular model can account for the detected quasi-periodic oscillations in our study.\par
The blazar's emission is mainly dominated by jets, making it probable that the observed quasi-periodic features are linked to the jet emissions. If the jet undergoes a precession, the QPO patterns could arise due to changes in the Lorentz factors along the observer's line of sight \citep{begelman1980massive, graham2015possible}.\par
Moreover, the jet orientation could also be influenced by the Lense-Thirring precession of the inner edge of the disc. However, these processes typically result in timescales on the order of years \citep{rieger2007supermassive}, considerably longer than the periods we report in our findings.\par
The jet-induced quasi-periodic oscillation involves a situation where a relativistic plasma travels in a helical path within the jet \citep{mohan2015kinematics}. These helical structures may form due to interactions with the surroundings or hydrodynamic instability. The observed periodic fluctuations are caused by changes in the Doppler boosting factor as the viewing angle of the plasma blob varies. Depending on factors like the Doppler boosting factor, pitch angle, and viewing angle, the variability timescale can range from a few days to several months. In this scenario, the blob emits $\gamma$-rays through processes like External Compton (EC) and Synchrotron Self Compton (SSC) in a one-zone leptonic setup. The blob's helical motion results in a changing viewing angle over time relative to our line of sight, given as \citep{sobacchi2016model, Zhou2018Nov, roy2022transient}
\begin{equation}
    \cos{\theta_{\text{obs}}(t)} = \sin{\phi}\sin{\psi}\cos{2\pi{t}/P_{\text{obs}}} + \cos{\phi}\cos{\psi}
\end{equation}
where $P_{obs}$ is the observed periodicity, $\phi$ is the pitch angle of the blob, and $\psi$ is the viewing angle or inclination angle measured between the observer's line of sight and the jet axis. The Doppler factor undergoes temporal variations, as described by $\delta = 1/\{\Gamma[1 - \beta cos\theta_{obs}(t)]\}$, where $\Gamma = 1/ \sqrt{1 - \beta^2}$, represents the bulk Lorentz factor associated with the motion of blob, and $\beta = \frac{v_{jet}}{c}$. Specifically, for an FSRQ, we consider typical values of $\phi = 2^{\circ}$, $\psi = 5^{\circ}$, and $\Gamma = 15$ \citep{Abdo2010May, sobacchi2016model, Zhou2018Nov}. The periods in the observed and rest frame of the blob can be defined as

\begin{equation}
    P_{rest} = \frac{P_{obs}}{1 - \beta cos\psi cos\phi}
\end{equation}

Using the expression of $\Gamma$, $\beta$ is estimated to be 0.9977 and the periodicity in the rest frame of the blob is estimated to be $P_{rest}$= 38.57 years for the $P_{obs}$= 93 days. The distance travelled by the blob in one period is $D_{1P} = c \beta$ $P_{rest}$ $cos\phi$ $\approx$ 11.79 pc. Parsec-scale jets have been identified in several blazars so far \citep{bahcall1995hubble, vicente1996monitoring, tateyama1998observations}, and the optical polarization observations have supported the existence of these helical structures \citep{marscher2008inner}, but their exact origins remain a mystery. To unravel the source of these helical features of the jets, a helical magnetic field could be a likely explanation \citep{vlahakis2004magnetic}. In the straight jet model, the inclination angle of the jet remains constant relative to the line of sight over time. However, in a modified helical jet model, the blob moves helically inside the curved jet \citep{sarkar2021multiwaveband, roy2022transient, Prince2023Oct}. In this setup, the inclination angle of the jet axis relative to the line of sight could be time-dependent, $\psi \equiv \psi(t)$ \citep{sarkar2021multiwaveband}. The variation in angle with time could explain why the amplitude of the oscillation changes with time. Observed flux modulation of PKS 2255-282 spanning over more than 2 years within the domain of investigation is likely caused by the movement of an enhanced emission region along the helical magnetic field within the curved jet. 

\section{ACKNOWLEDGMENT}
\begin{acknowledgments}
A. Sharma is grateful to Prof. Sakuntala Chatterjee at S.N. Bose National Centre for Basic Sciences, for providing the necessary support to conduct this research.
\end{acknowledgments}

%

\vspace{5mm}
\facilities{Fermi-LAT}








\bibliography{sample631}{}
\bibliographystyle{aasjournal}



\end{document}